\documentclass[12pt]{article}
\usepackage{amsmath}
\usepackage{latexsym}
\usepackage{amssymb}
\usepackage{a4wide}
\usepackage{bbm}
\usepackage{epsfig}
\newcommand{\I}{\text{i}}
\newcommand{\E}{\text{e}}

\newcommand{\STr}{\text{STr}}
\newcommand{\Tr}{\text{Tr}}
\newcommand{\be}{\begin{equation}}
\newcommand{\ee}{\end{equation}}
\newcommand{\bear}{\begin{eqnarray}}
\newcommand{\ear}{\end{eqnarray}}

\newcommand{\re}[1]{~(\ref{#1})}
\newcommand{\case}[2]{{\scriptstyle \frac{#1}{#2}}}
\newcommand{\Gk}{\Gamma_k}
\newcommand{\yl}{\psi_{\text{L}}}
\newcommand{\yr}{\psi_{\text{R}}}
\newcommand{\ybl}{\bar{\psi}_{\text{L}}}
\newcommand{\ybr}{\bar{\psi}_{\text{R}}}
\newcommand{\yb}{\bar{\psi}}
\newcommand{\pl}{P_{\text{L}}}
\newcommand{\pr}{P_{\text{R}}}
\newcommand{\lk}{\bar{\lambda}_{\sigma,k}}
\newcommand{\mkq}{\bar{m}_k^2}
\newcommand{\hk}{\bar{h}_k}
\newcommand{\Zphi}{Z_{\phi,k}}
\newcommand{\Zphik}{Z_{\phi,k}k^2}
\newcommand{\dlk}{\Delta\lk}
\newcommand{\hlk}{\hat{\lambda}_{\sigma,k}}
\newcommand{\hmkq}{\hat{m}_k^2}
\newcommand{\hhk}{\hat{h}_k}

\newcommand{\alb}{\hat{\alpha}_k}
\newcommand{\phih}{\hat{\phi}}
\newcommand{\Phih}{\hat{\Phi}}
\newcommand{\fss}[1]{#1\!\!\!/}
\newcommand{\fsl}[1]{#1\!\!\!\!/}
\newcommand{\SmP}{\ybr\yl\ybl\yr}
\newcommand{\Yc}{(\ybr\yl \phi-\ybl\yr\phi^\ast)}
\newcommand{\phid}{\phi^\ast}
\newcommand{\phihd}{\hat{\phi}^\ast}
\newcommand{\pat}{\partial_t}
\newcommand{\patt}{\tilde{\partial}_t}
\newcommand{\kB}{\Lambda}
\newcommand{\te}{\tilde{\epsilon}}

\newcommand{\talpha}{\tilde{\alpha}}
\begin{document}

$\text{}$

\vspace{-2cm}

{\hfill CERN-TH/2001-196}

{\hfill HD-THEP-01-31}

\vspace{1.5cm}

\centerline{\Large\bf Renormalization Flow of Bound States} 

\vspace{.8cm}

\centerline{\large Holger Gies${}^a$ and Christof Wetterich${}^b$}
 
\vspace{.6cm}

\centerline{\small\it ${}^a$ CERN, Theory Division, CH-1211 Geneva 23,
  Switzerland }
\centerline{\small\it \quad E-mail: Holger.Gies@cern.ch}

\vspace{.1cm}

\centerline{\small\it ${}^b$ Institut f\"ur theoretische Physik,
  Universit\"at Heidelberg,}
\centerline{\small\it Philosophenweg 16, D-69120 Heidelberg,
  Germany} 
\centerline{\small\it \quad E-mail: C.Wetterich@thphys.uni-heidelberg.de}

\begin{abstract}
  A renormalization group flow equation with a scale-dependent
  transformation of field variables gives a unified description of
  fundamental and composite degrees of freedom. In the context of 
  the effective average action, we study the renormalization flow of
  scalar bound states which are formed out of fundamental fermions.
  We use the gauged Nambu--Jona-Lasinio model at weak gauge coupling
  as an example. Thereby, the notions of bound state or fundamental
  particle become scale dependent, being classified by the fixed-point
  structure of the flow of effective couplings.
\end{abstract}

\section{Introduction}
Bound states, as opposed to fundamental particles, are commonly
thought of as derived quantities, in the sense that the properties of
positronium or atoms can be computed from the known electromagnetic
interactions of their constituents.  The conceptual separation between
bound states and fundamental particles is, however, not always so
obvious. As an example, it has been proposed that the Higgs scalar can
be viewed as a top-antitop bound state \cite{A}, with a compositeness
scale much above the characteristic scale of electroweak symmetry
breaking.  The mass of the bound state (and therefore the scale of
electroweak symmetry breaking) depends in this model on a free
parameter characterizing the strength of a four-fermion interaction.
For a bound-state mass or momentum near the compositeness scale
$\Lambda$, all the usual properties of bound states are visible. If
the Higgs boson mass is substantially smaller than $\Lambda$, however,
the bound state behaves like a fundamental particle for all practical
aspects relating to momentum scales sufficiently below the
compositeness scale.  Depending on the momentum scale, the particle
can therefore appear either as a typical bound state or a fundamental
particle. The scale dependence of the physical picture can be cast
into the language of the renormalization group (RG) by considering a
scale-dependent effective action. It should be possible to understand
the issues related to bound states or composite fields in this
context. In this paper we demonstrate how the effective behavior as
bound state or ``fundamental particle'' in dependence on a parameter
of the model can be understood within the exact renormalization group
equation for the effective average action \cite{Wetterich:1993yh}.

In strong interactions, bound states or composite fields play an
essential role in the dynamics at low momenta. In particular, scalar 
quark-antiquark bound states are responsible for chiral symmetry
breaking with the associated dynamics of the pions. Furthermore,
it has been proposed that the condensation of a color octet composite
field may lead to ``spontaneous breaking of color'' \cite{B} with a
successful phenomenology of the spectrum and interactions of the light
pseudoscalars, vector  mesons and baryons. In order to verify 
or falsify such a proposal and connect the parameters of an 
effective low-energy description to the fundamental parameters of 
QCD one needs a reliable connection between short and long distance
within the RG approach. In such a formalism it is convenient to
represent fundamental particles and bound states by fields on
equal footing. For quark-antiquark bound states this can be achieved
by partial bosonization. A first picture of the flow of
bound states in the exact RG approach has been developed in \cite{C}. 
A shortcoming of these initial proposals is the fact 
that the bosonization is typically performed at a fixed scale.
In a RG picture it would seem more appropriate that the
relation between the fields for composite and fundamental particles
becomes scale dependent. Furthermore, the simple observation that
a typical bound-state behavior should not lead to the same relevant
(or marginal) parameters as in the case 
of fundamental particles has not been very apparent so far. 

In this paper we propose a modified exact renormalization group
equation which copes with these issues. The field variables themselves
depend on the renormalization scale $k$. For this purpose we use
$k$-dependent nonlinear field transformations \cite{Wetterich:1996kf},
\cite{D}.  As a consequence, partial bosonization can be performed
continuously for all $k$. This yields a description where explicit
four-quark interactions which have the same structure as those
produced by the exchange of a bound state are absent for every scale
$k$. These interactions are then completely accounted for by the
exchange of composite fields.  We will demonstrate this approach in a
simple model, whereas the more formal aspects can be found in
appendices. As a result, we conclude that ``fundamental behavior'' is
related to a flow governed by an infrared unstable fixed point with
the appropriate relevant parameters. For the typical ``bound-state
behavior'' such a fixed point does not govern the flow. The
parameters characterizing the bound-state mass and interactions are
rather determined by an infrared attractive (partial) fixed point and
become therefore predictable as a function of the relevant or marginal
parameters characterizing masses and interactions of other
``fundamental fields''.  As a consequence, the notions of bound state
or fundamental particle become scale dependent, with a possible
crossover from one behavior to another.

As a simplified model sharing many features of electroweak or
strong interactions we consider the gauged NJL model \cite{E}
(with one flavor $N_F=1$), with action
\bear\label{AA} 
S&=&\int d^4x\bigl[\bar\psi i\gamma^\mu(\partial_\mu+\I e
A_\mu)\psi+2\lambda_{
  \text{NJL}}(\bar\psi_R\psi_L)(\bar\psi_L\psi_R)\nonumber\\ 
&&\qquad\qquad +\case{1}{4}F_{\mu\nu}F_{\mu\nu}+\case{1}{2\alpha}
(\partial_\mu A_\mu)^2\bigr].\ear
We consider here a small gauge coupling $e$. This model has two 
simple limits: For small $\lambda_{\text{NJL}}$ we recover
massless quantum electrodynamics (QED), whereas for large 
enough $\lambda_{\text{NJL}}$ one expects spontaneous chiral symmetry
breaking. The region of validity of perturbative
electrodynamics can be established by comparing $\lambda_{\text{NJL}}$ 
to the effective four-fermion interaction generated by box diagrams in
the limit of vanishing external momenta:
\begin{eqnarray}\label{BB}
\Delta{\cal L}_B&=&\frac{1}{4}\Delta\lambda(\bar\psi\gamma_\mu
\gamma_5\psi)(\bar\psi\gamma_\mu\gamma_5\psi)\nonumber\\
&=&\Delta\lambda\left[2(\bar\psi_L\psi_R)(\bar\psi_R\psi_L)
+\frac{1}{4}(\bar\psi\gamma_\mu \psi)(\bar\psi\gamma_\mu
\psi)\right].
\end{eqnarray}
Since the box diagrams are infrared divergent in the chiral limit
of vanishing electron mass, we have introduced a scale by implementing
an infrared cutoff $\sim k$ in the propagators such that\footnote{We
note that $\Delta\lambda$ does not depend on the gauge-fixing parameter
$\alpha$.}
\begin{equation}\label{CC}
\Delta\lambda=6e^4\int\frac{d^4q}{(2\pi)^4}[q^2(1+r_B(q))]^{-2}
[q^2(1+r_F(q))^2]^{-1}
=\frac{9}{16\pi^2}\frac{e^4}{k^2}.
\end{equation}
(The second equality holds for the particular cutoff functions
$r_B,r_F$ described in appendix E.) As long as perturbation theory
remains valid (small $e$), and
$\lambda_{\text{NJL}}\lesssim\Delta\lambda$, we do not expect the
four-fermion interaction $\sim \lambda_{\text{NJL}}$ to disturb
substantially the physics of massless QED. (In this case
$\lambda_{\text{NJL}}$ is an irrelevant parameter in the
renormalization group (RG) language.)

The spontaneous breaking of the chiral symmetry for
$\lambda_{\text{NJL}}>\lambda_{\text{c}}$ has been studied by a
variety of methods \cite{E,F,Reenders:2000bg,Jungnickel:1996fp}. For
strong four-fermion interactions the dominant physics can be described
by a Yukawa interaction with an effective composite scalar field. The
phase transition at $\lambda_{\text{NJL}}=\lambda_{\text{c}}$ is of
second order. In the vicinity of this transition the composite scalar
has all the properties usually attributed to a fundamental field. In
particular, its mass is governed by a relevant parameter. In this
paper we present a unified description of all these different features
in terms of flow equations for the effective average action.

\section{Flow equation for the gauged NJL model}
\label{flowequation}
Our starting point is the exact renormalization group  equation for the
scale-dependent effective action $\Gamma_k$ in the form
\cite{Wetterich:1993yh} 
\begin{equation}
\pat\Gk=\frac{1}{2} \STr \{\pat R_k(\Gamma^{(2)}_k+R_k)^{-1}\}.
\label{FE}
\end{equation}
The solution $\Gamma_k$ to this equation interpolates between its
boundary condition in the ultraviolet $\Gamma_\Lambda$, usually given by the classical
action, and the effective action $\Gamma_{k=0}$, representing the
generating functional of the 1PI Green's functions. This flow is
controlled by the to some extent arbitrary positive function
$R_k(q^2)$ that regulates the infrared fluctuations
at a scale $k$ and falls off quickly for $q^2>k^2$. Indeed, the insertion
$\pat R_k$ suppresses 
the contribution of modes with momenta
$q^2\gg k^2$. The operator $\pat$
represents a logarithmic
derivative  $\pat=k\frac{d}{dk}$. The heart of the flow equation is the
fluctuation matrix $\Gk^{(2)}$ that comprises second functional
derivatives of $\Gk$ with respect to all fields, and together with
$R_k$ it corresponds to the exact inverse propagator at a 
given scale $k$.
The (super-)trace runs over momenta and 
all internal indices including momenta and
provides appropriate minus signs for the fermionic sector.

For our study, we use the following simple truncation for the gauged
NJL model including the scalars arising from bosonization \cite{Jungnickel:1996fp}
(Hubbard--Stratonovich transformation): 
\begin{eqnarray}
\Gk&=&\int d^4x\biggl\{ \yb\I\fss{\partial}\psi +2\lk\,\SmP \nonumber\\
&&\qquad\qquad+Z_{\phi,k}\partial_\mu\phid\partial_\mu\phi+\mkq\,
\phid\phi +\hk\Yc  
  \nonumber\\
&&\qquad\qquad+\case{1}{4} F_{\mu\nu}F_{\mu\nu} +\case{1}{2\alpha}
 (\partial_\mu
A_\mu)^2 -e\yb\fsl{A}\psi\biggr\}. \label{1}
\end{eqnarray}
This truncation is sufficient for our purposes. For quantitative
estimates some of the simplifications could be improved in future
work. This concerns, in particular: setting the fermion and
gauge-field wave function renormalization constants to 1, reducing an
a priori arbitrary scalar potential to a pure mass term, skipping all
vector, axialvector, etc. channels of the four-fermion interaction as
well as all higher-order operators, neglecting the running of the
gauge coupling $e$ and dropping all higher-order derivative terms.
Especially the gauge sector is treated insufficiently, although this
is appropriate for small $e$; for simplicity, we use Feynman gauge,
$\alpha=1$.  The running of the scalar wave function renormalization
$Z_{\phi,k}$ will also not be studied explicitly; since $Z_{\phi,k}$
is zero for the bosonization of a point-like four-fermion interaction,
we shall assume that it remains small in the region of interest.

Nevertheless, the essential points of how fermionic interactions may
be translated into the scalar sector can be studied in this simple
truncation. Of course, the truncation is otherwise not supposed to
reveal all properties of the system even qualitatively; in particular,
the interesting aspects of the gauged NJL model at strong
coupling \cite{Leung:1986sn,Aoki:1997fh,Reenders:1999fz}
cannot be covered unless the scalar potential is generalized. 
 
The truncation\re{1} is related to the bosonized gauged NJL model, if
we impose the relation  
\begin{equation}
\lambda_{\text{NJL}}:=\frac{1}{2}\,
  \frac{\bar{h}_{\Lambda}^2}{\bar{m}_{\Lambda}^2} 
\label{1.6} 
\end{equation}
as a boundary condition at the bosonization scale $\Lambda$ and
$\bar{\lambda}_{\sigma,\Lambda}=0$, $Z_{\phi,\Lambda}=0$; it is this
bosonization scale $\Lambda$ that we consider as the ultraviolet starting
point of the flow.  In fact, the action (\ref{AA}) can be recovered
by solving the field equation of $\phi$ as functional  of $\psi,
\bar\psi$ and reinserting the solution into Eq. (\ref{1}).

Using the truncation\re{1}, the flow equation\re{FE} can be boiled
down to first-order coupled differential equations for the couplings
$\mkq$, $\hk$ and $\lk$. For this, we rewrite Eq.\re{FE} in the form
\begin{equation}
\pat\Gk=\frac{1}{2}\,\STr \,\patt\,\ln (\Gk^{(2)}+R_k), \label{FE2}
\end{equation}
where the symbol $\patt$ specifies a formal derivative that acts only
on the $k$ dependence of the cutoff function $R_k$. Let us specify the
elements of Eq.\re{FE} more precisely:
\begin{equation}
\bigl(\Gk^{(2)}\bigr)_{ab}:= \frac{\overrightarrow{\delta}}{\delta
  \Phi_a^T}\, \Gk\,\frac{\overleftarrow{\delta}}{\delta \Phi_b} , \quad
\Phi=\left(\begin{array}{c} A\\ \phi\\ \phid\\ \psi\\ \yb^T
\end{array}\right), \quad \Phi^T=(A^T,\phi,\phid,\psi^T,\yb).
\label{new4}
\end{equation}
Here $A\equiv A_\mu$ is understood as a column vector, and $A^T$
denotes its Lorentz transposed row vector. For spinors the
superscript $T$ characterizes transposed quantities in Dirac
space. The complex scalars $\phi$ and $\phid$ as well as the fermions
$\yb$ and $\psi$ are considered as independent, but transposed
quantities are not: e.g., $\Phi$ and $\Phi^T$ carry the same
information. 

Performing the functional differentiation, the fluctuation matrix can
be decomposed as
\begin{equation}
\Gamma_k^{(2)}+R_k ={\cal P}+{\cal F}, \label{5}
\end{equation}
where ${\cal F}$ contains all the field dependence and ${\cal P}$ the
propagators including the cutoff functions.
Their explicit representations are given in appendix B. Inserting
Eq.\re{5} into Eq.\re{FE2}, we can perform an  expansion
in the number of fields,
 \begin{eqnarray}
\pat\Gk\!&=&\!\frac{1}{2}\, \STr\,\patt\, \ln ({\cal P}+{\cal F})
  \label{11}\\ 
&=&\!\frac{1}{2}\,\STr\,\patt\,
      \left(\!\frac{1}{{\cal P}}{\cal F}\!\right)- 
   \frac{1}{4}\,\STr\,\patt\,\left(\!\frac{1}{{\cal P}}{\cal F}\!\right)^2 
\!+\frac{1}{6}\,\STr\,\patt\,\left(\!\frac{1}{{\cal P}}{\cal F}\!\right)^3 
\!-\frac{1}{8}\,\STr\,\patt\,\left(\!\frac{1}{{\cal P}}{\cal F}\!\right)^4
+\dots\,. \nonumber
\end{eqnarray}
The dots denote field-independent terms and terms beyond our
truncation. For our purposes it suffices to take the fields constant
in space.

The corresponding powers of $\frac{1}{{\cal P}}{\cal F}$ can be
computed by simple matrix multiplication and the (super-)traces can be
taken straightforwardly. This results in the following flow equations
for the desired couplings:
\begin{eqnarray}
\pat\mkq\equiv\, \beta_m&=&8 k^2 v_4 l^{(F)\,4}_{1}(0)\, \hk^2,
  \nonumber\\ 
\pat\hk\equiv\,\, \beta_h&=&-16 k^2 v_4 l^{(F)\,4}_{1}(0)\, \lk\hk
  -16 v_4 l^{(FB)\, 4}_{1,1}(0,0)\, e^2 \hk, \nonumber\\
\pat\lk\equiv\,\beta_{\lambda_\sigma}&=&-24 k^{-2} v_4
  l^{(FB)\,4}_{1,2}(0,0)\, e^4 
  - 32 v_4 l^{(FB)\, 4}_{1,1}(0,0)\,  e^2 \lk \nonumber\\
&&+8 v_4\, \frac{1}{Z_{\phi,k}} l^{(FB)\, 4}_{1,1}
  (0,\case{\bar m^2_k}{Z_{\phi,k} k^2})\, \hk^2\lk  
  - 8 k^2 v_4 l^{(F)\,4}_1(0)\,  \lk^2\nonumber\\
&&+\frac{2v_4}{Z^2_{\phi,k}k^2}l^{(FB)4}_{1,2}\left(0,
\case{\bar m_k^2}{Z_{\phi,k}k^2}\right)\bar h_k^4, \label{23} 
\end{eqnarray}
where $v_4=1/(32\pi^2)$. The threshold functions $l$ fall off for
large arguments and describe the decoupling of particles with mass
larger than $k$. They are defined in \cite{Jungnickel:1996fp};
explicit examples are given in appendix \ref{optimized}. At this
point, it is important to stress that all vector (V) and axialvector
(A) four-fermion couplings on the right-hand side of the flow equation
have been brought into the form (V) and $(V+A)$, and then the $(V+A)$
terms have been Fierz transformed into the chirally invariant scalar
four-fermion coupling $(S-P)$ used in our truncation (cf. Eq.
(\ref{BB})).  The pure vector coupling is omitted for the time being
and will be discussed in Sect.~\ref{gaugefield}.  It should also be
mentioned that no tensor four-fermion coupling is generated on the
right-hand side of the flow equation.

Incidentally, the mass equation coincides with the results of
\cite{Jungnickel:1996fp}; we find agreement of the third
equation  with the results of \cite{Aoki:1997fh} where the same model
was investigated in a nonbosonized version.\footnote{Remaining
  numerical differences arise from wave function renormalization which
  was included in \cite{Aoki:1997fh}, but is neglected here for
  simplicity.}. 
We note that the last term in $\beta_{\lambda_
\sigma}$, which is $\sim \bar h_k^4$, is suppressed by the threshold
function as long as $\bar m^2/(Z_\phi k^2)$ remains large. For
simplicity of the discussion we will first omit it and
comment on its quantitative impact later on. The inclusion of this
term does not change the qualitative behavior.

\section{Fermion-boson translation by hand}
\label{byhand}

As mentioned above, the boundary conditions for the flow equation are
such that the four-fermion interaction vanishes at the bosonization
scale, $\bar{\lambda}_{\sigma,\Lambda}=0$. But lowering $k$ a bit
introduces the four-fermion interaction again according to Eq.\re{23}:
\begin{equation}
\pat\lk\bigr|_{k=\Lambda} =-24\Lambda^{-2} v_4
l^{(FB)\,4}_{1,2}(0,0)\, e^4 \neq 0. \label{2.1}
\end{equation}
In Eq.\re{1}, we may again solve the field equations for $\phi$ as a
functional of $\yb,\psi$ and find in Fourier space
\begin{equation}
\phi(q)=\frac{\hk(\ybl\yr)(q)}{\mkq+Z_{\phi,k}q^2}, \quad
\phid(q)=-\frac{\hk(\ybr\yl)(-q)}{\mkq+Z_{\phi,k}q^2}. \label{CW1}
\end{equation}
Inserting this result into $\Gamma_k$ yields the ``total'' four-fermion
interaction $(\int_q\equiv\int\left(\case{dq}{2\pi}\right)^4)$
\begin{equation}
\int_q \left( 2\lk+\frac{\hk^2}{\mkq+Z_{\phi,k}q^2}\right)
(\ybr\yl)(-q) (\ybl\yr)(q). \label{CW2}
\end{equation}
The local component (for $q^2=0$) contains a direct contribution $\sim
2\lk$ (one-particle irreducible in the bosonized version) and a scalar
exchange contribution (one-particle reducible in the bosonized
version). From the point of view of the original fermionic theory,
there is no distinction between the two contributions (both are 1PI in
the purely fermionic language). This shows a redundancy in our
parametrization, since we may change $\lk$, $\hk$ and $\mkq$ while
keeping the effective coupling
\begin{equation}
2\lambda_\sigma^{\text{eff}}(q)=2\lk+\frac{\hk^2}{\mkq+Z_{\phi,k}q^2}
\label{CW3}
\end{equation}
fixed. Indeed, a choice $\lk'$, $\hk'$, $\mkq{}'$ leads to the same
$\lambda_\sigma^{\text{eff}}(0)$ if it obeys
\begin{equation}
\lk'=\lk+\frac{\hk^2}{2\mkq} -\frac{\hk^2{}'}{2\mkq{}'}. \label{CW4}
\end{equation}
In particular, we will choose a parametrization where $\lk'$ vanishes
for all $k$. In this parametrization, any increase $d\lk$ according to
Eq.\re{23} is compensated for by a change of $\frac{1}{2}d\left(
  \frac{\hk^2}{\mkq}\right)$ of the same size. An increase in $\lk$ is
mapped into an increase in $\hk^2/(2\mkq)$. In this parametrization,
the four-fermion coupling remains zero, whereas the flow of $\hk^2/\mkq$
receives an additional contribution
\begin{equation}
\pat\left( \frac{\hk^2}{\mkq} \right)
=\pat\left( \frac{\hk^2}{\mkq} \right)\biggr|_{\lk}
  +2\pat\lk\biggr|_{\hk^2,\mkq}. \label{CW5}
\end{equation}
More explicitly, we can write
\begin{equation}
\pat\left( \frac{\bar m^2_k}{\bar h_k^2}\right)=\frac{1}{\hk^2} \pat
  \mkq\bigl|_{\lk} 
-2\frac{\mkq}{\hk^3} \pat\hk\bigl|_{\lk}
-2\frac{\bar{m}_k^4}{\hk^4}\pat\lk,
\label{CW6}
\end{equation}
with $\pat\mkq\bigl|_{\lk}$, $\pat\hk\bigl|_{\lk}$, $\pat\lk$ given by
Eq.\re{23} with the replacement $\lk\to 0$ on the right-hand sides.
One obtains
\begin{equation}
\pat\left(\frac{\mkq}{\hk^2} \right)=v_4\left[ 8\,l_1^{(F)\,4}(0)\, k^2
  +{32}\, l_{1,1}^{(FB)\, 4}(0,0)\, e^2\frac{\mkq}{\hk^2}
  +48\,l_{1,2}^{(FB)\, 4}(0,0)\, \frac{e^4}{k^2} 
    \left( \frac{\mkq}{\hk^2} \right)^2 \right]. \label{CW7}
\end{equation}

\begin{figure}
\begin{center}
\begin{picture}(80,45)
\put(0,3){
\epsfig{figure=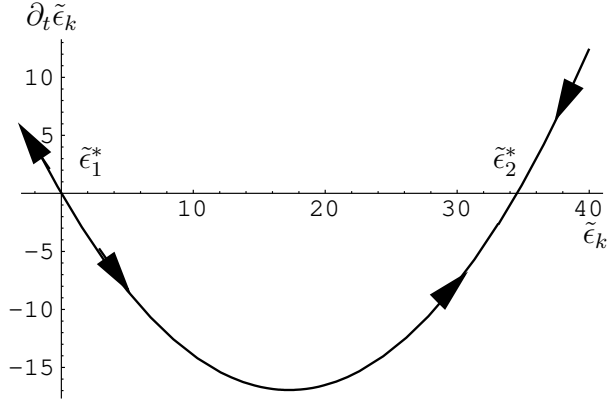,width=8cm}}
\put(4,53){$\pat \te_k$}
\put(78,24){$\te_k$} 
\put(11,34){$\te_1^\ast$} 
\put(66,34){$\te_2^\ast$}
\end{picture} 
\end{center}
\caption{Fixed-point structure of the $\te_k$ flow equation after 
  fermion-boson translation by hand. The graph is plotted for the
  threshold functions discussed in appendix \ref{optimized} with
  $e=1$. Note that $\te_1^\ast$ is small but different from zero
  (cf.~Eq.\re{O.11}). Arrows indicate the flow towards the infrared,
  $k\to 0$.}
\label{fig1} 
\end{figure}

The characteristics of this flow can
be understood best in terms of the  dimensionless quantity
\begin{equation}
\te_k=\frac{\mkq}{\hk^2 k^2}. \label{CW8}
\end{equation}
It obeys the flow equation
\begin{eqnarray}
\pat \te_k=\beta_{\te}\!&=&\!-2\te_k +v_4\left[ 8l_1^{(F)\,4}(0)+{32}
  l_{1,1}^{(FB)\, 4}(0,0)e^2\te_k +48 l_{1,2}^{(FB)\, 4}(0,0) e^4\te_k^2
  \right] \label{CW9}\\
&=&\!-2\te_k + \frac{1}{8\pi^2} +\frac{1}{\pi^2} e^2\, \te_k+
  \frac{9}{4\pi^2} e^4\, \te_k^2, \nonumber
\end{eqnarray}
where in the last line we have inserted the values of the threshold
functions for optimized cutoffs \cite{Litim:2001up} discussed in
appendix \ref{optimized}. Neglecting the running of the gauge coupling
$e$, we note in Fig.~\ref{fig1} the appearance of two fixed points.
For gauge couplings of order 1 or smaller and to leading order in $e$,
these two fixed points are given by
\begin{equation}
\te_1^\ast\simeq \frac{1}{16\pi^2} +{\cal O}(e^2/(16\pi^2)^2), \quad
\te_2^\ast\simeq \frac{8\pi^2}{9e^4} +{\cal O}(1/e^2). \label{O.11}
\end{equation}
 
The smaller fixed point $\te^\ast_1$ is infrared unstable, whereas the
larger fixed point $\te^\ast_2$ is infrared stable. Therefore,
starting with an initial value of $0<{\te}_\Lambda<\te^\ast_1$, the
flow of the scalar mass-to-Yukawa-coupling ratio will be dominated by
the first two terms in the modified flow equation\re{CW9} $\sim -2
\te_k+1/(8\pi^2)$. This is nothing but the flow of a theory involving
a ``fundamental'' scalar with Yukawa coupling to a fermion sector.
Moreover, we will end in a phase with (dynamical) chiral symmetry
breaking, since $\tilde\epsilon$ is driven to negative values. (Higher
order terms in the scalar potential need to be included once
$\tilde\epsilon$ becomes zero or negative.)  This all agrees with the
common knowledge that the low-energy degrees of freedom of the strongly
coupled NJL model are (composite) scalars which nevertheless behave as
fundamental particles.

On the other hand, if we start with an initial $\te_\Lambda$ value
that is larger than the first (infrared unstable) fixed point, the
flow will necessarily be attracted towards the second fixed point
$\te^\ast_2$; there, the flow will stop. This flow does not at all
remind us of the flow of a fundamental scalar. Moreover, there will be
no dynamical symmetry breaking, since the mass remains positive.  The
effective four-fermion interaction corresponding to the second fixed
point reads
\begin{equation}\label{19a}
\lambda^*_\sigma=\frac{1}{2k^2\tilde\epsilon^*_2}\approx
\frac{9}{16\pi^2} \frac{e^4}{k^2}.
\end{equation}
It coincides with the perturbative value (3) of massless QED. We
conclude that the second fixed point characterizes massless QED.  The
scalar field shows a typical bound-state behavior with mass and
couplings expressed by $e$ and $k$. (The question as to whether the
bound state behaves like a propagating particle (i.e., ``positronium'')
depends on the existence of an appropriate pole in the scalar
propagator. At least for massive QED one would expect such a pole with
renormalized mass corresponding to the ``rest mass'' of scalar
positronium.)

From a different viewpoint, the fixed point $\te^\ast_1$ corresponds
directly to the critical coupling of the NJL model, which
distinguishes between the symmetric and the broken phase. As long as
the flow is governed by the vicinity of this fixed point, the scalar
behaves like a fundamental particle, with mass given by the relevant
parameter characterizing the flow away from this fixed point.

Our interpretation is that the ``range of relevance'' of these two
fixed points tell us whether the scalar appears as a ``fundamental''
or a ``composite'' particle, corresponding to the state of the system
being governed by $\te^\ast_1$ or $\te^\ast_2$, respectively.

The incorporation of the flow of the momentum-independent part of
$\lk$ into the flow of $\hk$ and $\mkq$ affects only the ratio
$\mkq/\hk^2$. At this point, it does not differentiate which part of
the correction appears in the separate flow equations for $\mkq$ and
$\hk$, respectively. This degeneracy can be lifted if we include
information about the flow of $\lk$ for two different values of the
external momenta. Let us define $\lk(s)$ as $\lk(p_1,p_2,p_3,p_4)$
with $p_1=p_3=(1/2)(\sqrt{s},\sqrt{s},0,0)$,
$p_2=p_4=(1/2)(\sqrt{s},-\sqrt{s},0,0)$, where
$s=(p_1+p_2)^2=(p_3+p_4)^2$ is the square of the exchanged momentum in
the $s$ channel \cite{Wetterich:1996kf}. The coupling $\lk$ appearing
on the right-hand side of Eq.\re{CW5} corresponds in this notation to
$\lk(s=0)$. We can now achieve the simultaneous vanishing of
$\lk(s=0)$ and $\lk(s=k^2)$ if we redefine $\mkq$ and $\hk$ such that
they obey in addition
\begin{equation}
\pat \left( \frac{\hk^2}{\mkq+\Zphik}\right)=
\pat \left( \frac{\hk^2}{\mkq+\Zphik}\right)_{|\lk} +2\pat
\lk(s=k^2). \label{23a}
\end{equation}
Incorporation of this effect should improve a truncation where the
1PI four-fermion coupling is neglected subsequently, since we realize
now a matching at two different momenta.

The combination of Eqs.\re{CW5} and\re{23a} specifies the evolution of
$\mkq$ and $\hk$,
\begin{eqnarray}
\pat\mkq&=&\beta_m\bigl|_{\lk}+ \frac{2\mkq(\mkq+\Zphik)}{\hk^2}
\left( \frac{\mkq+\Zphik}{\Zphik}\, \pat\dlk+\pat\lk(s=0)\right),
\label{23b}\\
\pat\hk&=&\beta_h\bigl|_{\lk} + \frac{2\mkq +\Zphik}{\hk}\,
\pat\lk(s=0) + \frac{(\mkq+\Zphik)^2}{\Zphik\, \hk}\, \pat\dlk,
\label{23c}
\end{eqnarray}
where we have used 
\begin{equation}
\dlk=\lk(s=k^2)-\lk(s=0). \label{23d}
\end{equation}
%

Let us finally comment on the influence of the last term $\sim
\bar{h}_k^4$ of Eq.\re{23}, omitted up to now, on the flow
equation\re{CW9} for $\te_k$: the contribution of this term to
Eq.\re{CW9} is $\sim (\case{\mkq}{Z_{\phi,k}k^2})^2 \, l_{1,2}^{(FB)\,
  4} (0,\case{\mkq}{Z_{\phi,k}k^2})$. For large
$(\case{\mkq}{Z_{\phi,k}k^2})$, this term approaches a constant, so
that a slight vertical shift of the parabola of Fig.~\ref{fig1} is
induced. We observe that this shift leaves the position of the second
fixed point $\te^\ast_2$ unaffected to lowest order in $e$.  This
justifies the omission of the $\sim \bar{h}_k^4$ term in the preceding
discussion. The influence of the $\hk^4$ term on the first fixed point
is discussed at the end of the next section.

\section{Flow with continuous fermion-boson translation}
\label{conti}

The ideas of the preceding section shall now be made more rigorous by
deriving the results directly from an appropriate exact flow equation.
As a natural approach to this aim, we could search for a $k$-dependent
field transformation of the scalars, $\phi\to\phih_k[\phi]$.  In terms
of the new variables, the flow equation (\ref{FE2}) should then
provide for the vanishing of the four-fermion coupling in the
transformed effective action. Indeed, we sketch this approach briefly
in appendix \ref{fieldtrafo}. Instead, we propose here a somewhat
different approach relying on a variant of the usual flow equation
where the cutoff is adapted to $k$-dependent fields. The advantage is
a very simple structure of the resulting flow equations in coincidence
with those of the preceding section.

The idea is to employ a flow equation for a scale-dependent effective
action $\Gamma_k[\phi_k]$, where the field variable $\phi_k$ is
allowed to vary during the flow; we derive this flow equation in
appendix \ref{flowvar}. To be precise within the present context, upon
an infinitesimal renormalization group step from a scale $k$ to
$k-dk$, the scalar field variables also undergo an infinitesimal
transformation of the type (in momentum space)
\begin{eqnarray}
\phi_{k-dk}(q)&=&\phi_k(q)+\delta\alpha_k(q)\, (\ybl\yr)(q)
  \nonumber\\
&\equiv&\phi_k(q)+\delta\alpha_k(q)\int_p\,
  \ybl(p)\yr(p+q). \label{ffv13}
\end{eqnarray}
Including the corresponding
transformation of the complex conjugate variable, the flow of the
scalar fields is given by
\begin{eqnarray}
\pat\phi_k(q)&=&-(\ybl\yr)(q)\, \pat\alpha_k(q), \nonumber\\
\pat\phid_k(q)&=&(\ybr\yl)(-q)\, \pat\alpha_k(q).\label{d13b}
\end{eqnarray}
The  transformation parameter $\alpha_k(q)$ is an a 
priori arbitrary function, expressing a redundancy in the
parametrization of the effective action. As shown in Eq.\re{ffv12},
the effective action $\Gamma_k[\phi_k,\phid_k]$ obeys the modified
flow equation
\begin{equation}
\pat\Gamma_k[\phi_k,\phid_k]=
  \pat\Gamma_k[\phi_k,\phid_k]\bigl|_{\phi_k,\phid_k}
  +\int_q\left( \frac{\delta\Gamma_k}{\delta\phi_k(q)}\, \pat\phi_k
  (q) + \frac{\delta\Gamma_k}{\delta\phid_k(q)}\, \pat\phid_k(q)\right),
  \label{ffv16} 
\end{equation}
where the notation omits the remaining fermion and gauge fields for
simplicity. The first term on the right-hand side is nothing but the
flow equation for fixed variables evaluated at $\phi_k$, $\phid_k$
instead of $\phi,\phid=\phi_\Lambda,\phid_\Lambda$. 
The second term reflects the flow of the variable. Projecting
Eq.\re{ffv16} onto our truncation\re{1}, we arrive at modified flows
for the couplings: 
\begin{eqnarray}
\pat\mkq&=&\pat\mkq\bigl|_{\phi_k,\phid_k}, \nonumber\\
\pat\hk&=&\pat\hk\bigl|_{\phi_k,\phid_k}+(\mkq+Z_{\phi,k}q^2)\,
  \pat\alpha_k(q), \label{ffv17}\\
\pat\lk&=&\pat\lk\bigl|_{\phi_k,\phid_k}-\,\hk\,
\pat \alpha_k(q). \nonumber
\end{eqnarray}
Again, the first terms on the right-hand sides are nothing but the
right-hand sides of Eq.\re{23}, i.e., the corresponding beta functions
$\beta_{m,h,\lambda_\sigma}$. The further terms represent the
modifications owing to the flow of the field variables, as obtained
from the two last terms in Eq. (\ref{ffv16}) by inserting Eq.
(\ref{d13b}). Obviously, we could have generalized the method easily
to the case of momentum-dependent couplings (see below). In the
following, however, it suffices to study the point-like limit, which
we associate to $q=0$.

We exploit the freedom in the choice
of variables in Eq.\re{d13b} by
fixing $\alpha_k=\alpha_k(q=0)$ in such a way that the
four-fermion coupling is not renormalized, $\pat\lk=0$. This implies
the flow equation for $\alpha_k$,
\begin{equation}
\pat\alpha_k=\beta_{\lambda_\sigma}/\bar h_k. \label{ffv18}
\end{equation}
Together with the boundary condition $\lambda_{\sigma,\Lambda}=0$,
this guarantees a vanishing four-fermion coupling at all scales,
$\lk=0$. The (nonlinear) fields corresponding to this choice
obtain for every $k$ by integrating the flow (\ref{ffv18})
for $\alpha_k$, with $\alpha_\Lambda=0$.

Of course, imposing the condition\re{ffv18} also influences the
flow of the Yukawa coupling according to Eq.\re{ffv17},
\begin{equation}
\pat\hk=\beta_h+\frac{\mkq}{\hk}\, \beta_{\lambda_\sigma}.
\label{ffv19}
\end{equation}
In consequence, the flow equation for the quantity of
interest, $\hk^2/\mkq$, then reads
\begin{equation}
\pat\left(\frac{\hk^2}{\mkq}\right)
=\pat\left(\frac{\hk^2}{\mkq}\right)\biggl|_{\phi_k,\phid_k}
+2\beta_{\lambda_\sigma}. \label{ffv20}
\end{equation}
This coincides precisely with Eq.\re{CW5} where we have translated the
fermionic interaction into the scalar sector by hand.  The flow
equation of the dimensionless combination
$\te_k=\frac{\mkq}{k^2\hk^2}$ is therefore identical to the one
derived in Eq.\re{CW9}, so that the fixed-point structure described
above is also recovered in the more rigorous approach. The underlying
picture of this approach can be described as a permanent translation
of four-fermion interactions, arising during each renormalization
group step, into the scalar interactions. Thereby, bosonization takes
place at any scale and not only at a fixed initial one.

One should note that the field transformation is not fixed uniquely
by the vanishing of $\bar\lambda_{\sigma,k}$. For instance,
an additional contribution in Eq. (\ref{ffv13}) $\sim\delta
\beta_k(q)\phi_k(q)$ can absorb the momentum dependence of
the Yukawa coupling by modifying the scalar propagator. Similarly to
the discussion in Sect.~\ref{byhand}, this can be used in order to
achieve simultaneously the vanishing of $\lk(s)$ for all $s$ and $k$
and a momentum-independent $\hk$. First, the variable change
\begin{eqnarray}
\pat\phi_k(q)&=&-(\ybl\yr)(q)\, \pat\alpha_k(q)+ \phi_k(q)\,
\pat\beta_k(q), \nonumber\\
\pat\phid_k(q)&=&(\ybr\yl)(-q)\, \pat\alpha_k(q)+ \phid_k(q)\,
\pat\beta_k(q)\label{d13ba}
\end{eqnarray}
indeed ensures the vanishing of $\lk(s=q^2)$ if
$\pat\alpha_k(q)=\hk^{-1} \pat\lk(q^2)$. This choice results in 
\begin{eqnarray}
\pat\hk(q)&=&\pat\hk(q)\bigl|_{\lk} + \frac{\Zphi q^2+\mkq}{\hk}\,
\pat\lk(q^2) +\hk\,\pat\beta_k(q), \nonumber\\
\pat\Zphi(q) q^2+\pat \mkq&=&\pat\mkq\bigl|_{\lk}
+2\pat\beta_k(q)\,(\Zphi q^2+\mkq), \label{30a}
\end{eqnarray}
where $\hk(q)$ and $\Zphi(q)$ depend now on $q^2$. Secondly, the
momentum dependence of $\hk(q)$ can be absorbed by the choice
\begin{equation}
\pat\beta_k(q)=-\frac{\Zphi q^2+\mkq}{\hk^2}\, \pat\lk(q^2)
   + \frac{1}{\Zphik \hk^2} \left[ (\Zphik +\mkq)^2\pat\lk(k^2)
   -\bar{m}_k^4\pat\lk(0) \right]. \label{30b}
\end{equation}
The particular form of the $q$-independent part of $\pat\beta_k$ was
selected in order to obtain $\pat \Zphi(q^2=k^2)=0$ such that our
approximation of constant $\Zphi$ is self-consistent. Inserting
Eq.\re{30b} into the evolution equation\re{30a} for $\hk$ and $\mkq$, we
recover Eqs.\re{23b} and\re{23c}. We also note that the evolution of
$\te=\mkq/(\hk^2(0)k^2)$ is independent of the choice of $\beta_k(q)$.

It is interesting to observe that reinserting the classical equations
of motion at a given scale in order to eliminate auxiliary variables
is equivalent to the here-proposed variant of the flow equation with
flowing variables. In contrast, the standard form of the flow equation
in combination with a  variable transformation, to be discussed
in appendix \ref{fieldtrafo}, leads to a more complex structure, which
is in general more difficult to solve.

\section{Between massless QED and \\
  spontaneous chiral symmetry breaking}
\label{rencoup}

In this section, we briefly present some quantitative results for the
flow in the gauged NJL model. Despite our rough approximation, they
represent the characteristic physics. We concentrate on the flow of
the dimensionless renormalized couplings 
\begin{equation}
\epsilon_k=\frac{\mkq}{Z_{\phi,k}k^2}, \quad
h_k=\hk \, Z_{\phi,k}^{-1/2}, \quad 
\te_k=\frac{\epsilon_k}{h_k^2}, \quad
\talpha_k=\alpha_k\, Z_{\phi,k}^{1/2}\, k^2 \label{A}
\end{equation}
in the symmetric phase. Inserting the specific
threshold functions of appendix \ref{optimized}, we find the system of
differential flow equations
\begin{eqnarray} 
\pat \epsilon_k&=& -2 \epsilon_k + \frac{h_k^2}{8\pi^2}
-\frac{\epsilon_k(\epsilon_k+1)}{h_k^2} \left(\frac{9e^4}{4\pi^2}
  -\frac{h_k^4}{16\pi^2} \frac{3+\epsilon_k}{(1+\epsilon_k)^3} \right)
\bigl(1+(1+\epsilon_k)Q_\sigma\bigr), \nonumber\\
\pat h_k&=& -\frac{e^2}{2\pi^2}\, h_k
-\frac{2\epsilon_k+1+(1+\epsilon_k)^2Q_\sigma}{h_k}\left(
  \frac{9e^4}{8\pi^2} -\frac{h_k^4}{32\pi^2}
  \frac{3+\epsilon_k}{(1+\epsilon_k)^3} \right). \label{B}  
\end{eqnarray}
The resulting flow for $\te_k$ is independent of $Q_\sigma\equiv
\pat\dlk/\pat\lk(0)$:
\begin{equation}
\pat \te_k=\beta_{\te}=-2\te_k + \frac{1}{8\pi^2} 
  +\frac{e^2}{\pi^2} \te_k + \frac{9e^4}{4\pi^2} \te_k^2
  -\frac{1}{16\pi^2}\frac{\epsilon_k^2(3+\epsilon_k)}{(1+\epsilon_k)^3}.
  \label{C} 
\end{equation}
Here the last term reflects the last contribution to
$\beta_{\lambda_\sigma}$ in Eq.\re{23}, which has been neglected in
the preceding section (cf. Eq.\re{CW9}). We see that its influence is
small for $\epsilon_k\ll 1$, whereas for $\epsilon_k\gg1$ it reduces
the constant term by a factor $1/2$.

Note that, for a given $Q_\sigma$, Eqs.\re{B} form a closed set of
equations. The same is true for the flows of $\te_k$ and $\epsilon_k$
if we express $h_k$ in terms of $\te_k$ and $\epsilon_k$ in the first
line of Eq.\re{B}. In order to obtain $Q_\sigma$, the flow of $\lk(s)$
has to be known; however, far less information is already sufficient
for a qualitative analysis. First, it is natural to expect that
$\lk(s)$ is maximal for $s=0$, since $\lk(s)$ will be suppressed for
large $s$ owing to the external momenta. This implies
$\dlk/\lk(s=0)<0$. With the simplifying assumption that
$\dlk/\lk(0)\simeq$ const., we also conclude that
\begin{equation}
Q_\sigma<0.\label{Qsigma}
\end{equation}
For a qualitative solution of the flow equations, we assume
$|Q_\sigma|$ to be of order 1 or smaller. 

We next need initial values $\epsilon_\Lambda$, $\te_\Lambda$ for
solving the system of differential equations. We note that the initial
value $\epsilon_\Lambda$ diverges for the pure NJL model, since
$Z_{\phi,\Lambda}=0$. For large $\epsilon_k$, one has
\begin{equation}
\pat \epsilon_k =\left[-2+ \frac{1}{8\pi^2 \te_k} 
  +(|Q_\sigma|\, \epsilon_k-1) \left(\frac{9e^4\te_k}{4\pi^2}
  -\frac{1}{16\pi^2 \te_k}\right) \right] \epsilon_k, \label{D}
\end{equation}
and we find that $\epsilon_k$ decreases rapidly for
\begin{equation}
\te_\Lambda>\frac{1}{6e^2}. \label{tebound}
\end{equation}
(In a more complete treatment, it decreases rapidly for arbitrary
$\te_k$ owing to the generation of a nonvanishing $Z_{\phi,k}$ by the
fluctuations. In the present truncation, the qualitative behaviour of
the flow will depend on the details of $Q_\sigma$ if $\te_\Lambda$
does not satisfy this bound.)  We confine our discussion to initial
values satisfying Eq.\re{tebound}, which can always be accomplished
without fine-tuning.

\begin{figure}
\begin{picture}(160,45)
\put(0,0){
\epsfig{figure=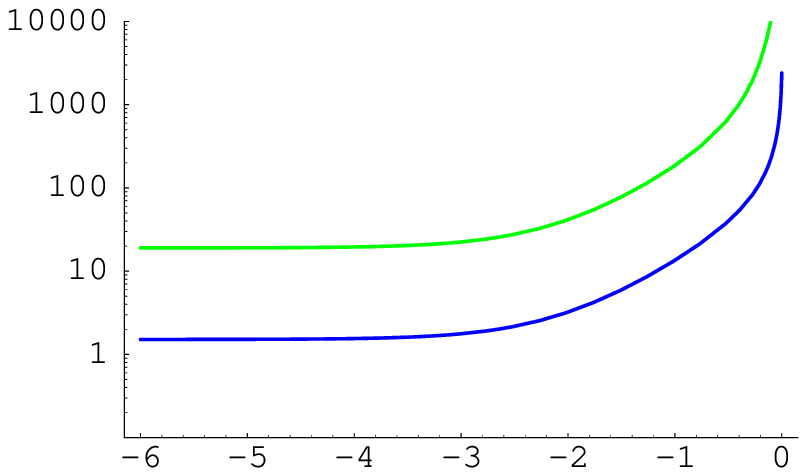,width=7.5cm}}
\put(66,38){$\epsilon_k$}
\put(70,21){$h_k$}
\put(69,0){$t$}
\put(84,0){ 
\epsfig{figure=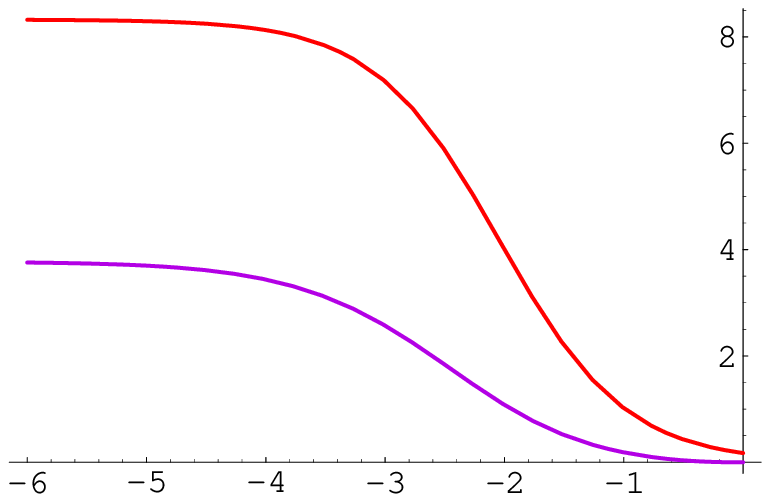,width=7.3cm}}
\put(95,40){$\te_k$}
\put(93,23){$100\cdot\talpha_k$}
\put(89,6){$t$}
\end{picture} 
\caption{Flows of $\epsilon_k$, $h_k$, $\te_k$ and $\talpha_k$ in the
  symmetric phase according to Eqs.\re{B},\re{C} and\re{E}
  for the initial values $\epsilon_{\kB}=10^6$,
  $\te_\Lambda=0.17\gtrsim 1/(6e^2)$, $e=1$, $Q_\sigma=-0.1$. For a
  better visualization, $\talpha_k$ has been multiplied by a factor of
  100. The plot of $\te_k$ on the right panel exhibits the crossover
  behavior between the fixed points $\te_1^\ast$ at small $t$ to
  $\te_2^\ast$ for $t\to -\infty$.}
\label{figXX} 
\end{figure}

In Fig.~\ref{figXX} we present a numerical
solution for large $\epsilon_\Lambda$ (small nonzero $Z_{\phi,k}$),
$Q_\sigma=-0.1$, and $\te_\Lambda$ slightly above the bound given by
Eq.\re{tebound} for $e=1$; these initial conditions correspond to
the symmetric phase. We observe that both $h_k$ and 
$\epsilon_k$ approach constant values in the infrared. This
corresponds to the ``bound-state fixed point'' for $\te_k$:
$\te_2^\ast\simeq 8\pi^2/(9e^4)$. A constant
$\epsilon_k$ implies that the renormalized mass term $m_k^2=\epsilon_k
k^2$ decreases $\sim k^2$ in the symmetric phase. The precise value of
the Yukawa coupling at the fixed point depends on $e$ and $|Q_\sigma|$:
\begin{equation}
(h^\ast)^2=16\pi^2\epsilon^\ast -
\frac{8\epsilon^\ast(\epsilon^\ast+1)(1-|Q_\sigma|(\epsilon^\ast+1))
  }{(2\epsilon^\ast+1-|Q_\sigma|(\epsilon^\ast+1)^2)} e^2. \label{34a}
\end{equation}
If $\epsilon^\ast\gg1 $ still holds, the fixed-point values can be
given more explicitly:
\begin{equation}
\epsilon^\ast\simeq \frac{2}{|Q_\sigma|}, \quad
h^\ast\simeq\frac{3e^2}{2\pi\sqrt{|Q_\sigma|}}. \label{fixpexp}
\end{equation}
Note that $\epsilon^\ast\gg1 $ is equivalent to $|Q_\sigma|\ll 1$;
numerically, we find that Eqs.\re{fixpexp} describe the fixed-point
values reasonably well already for $|Q_\sigma|\lesssim0.1$. We
observe that the fixed-point values are independent of the initial
values $\epsilon_\Lambda$ and $\te_\Lambda$, so that the system has
``lost its memory''.  

Finally, the parameter $\talpha_k$ governing the field redefinition
obeys the flow equation
\begin{equation}
\pat \talpha_k=2\talpha_k-\frac{9e^4}{8\pi^2\, h_k} +
\frac{h_k^3}{32\pi^2}\, \frac{3+\epsilon_k}{(1+\epsilon_k)^3}. \label{E}
\end{equation}
A numerical solution is plotted in Fig.~\ref{figXX}, right panel. Also
$\tilde{\alpha}_k$ approaches a constant for small $k$. Therefore, the
transformation parameter $\alpha_k\sim \tilde{\alpha}_k/k^2$ increases
for small $k$.

The physical picture of the fixed point\re{fixpexp} is quite
simple. We may first translate back to an effective four-fermion
interaction by solving the scalar field equations:
\begin{equation}
\lk(q^2)=\frac{1}{2} \frac{ (h^\ast)^2}{(q^2+\epsilon^\ast\, k^2)} 
  =\frac{9e^4}{8\pi^2} \frac{1}{(|Q_\sigma| q^2 +2
  k^2)}. \label{transback}
\end{equation}
In the limit $k\to0$, this mimics the exchange of a massless
positronium-like state with effective coupling $h^\ast=3e^2/(2\pi
\sqrt{|Q_\sigma|})$. Indeed, if we switch on the electron mass
$m_{\text{e}}$, we expect that the running of the positronium mass
term stops at $k^2\simeq m_{\text{e}}^2$. In consequence, the
positronium state will acquire a mass $\sim m_{\text{e}}$, which is,
in principle, calculable by an improved truncation within our
framework. 

On the other hand, starting with small enough $\te_\Lambda$, one will
observe chiral symmetry breaking as we have already argued in
Sect.\re{byhand}. Quantitative accuracy should include at least the
flow of the scalar wave function renormalization in this case. 
Near the boundary between the two phases, the infrared physics is
described by a renormalizable theory for QED with a neutral scalar
coupled to the fermion.

\section{Modified gauge fields}
\label{gaugefield}

The possibility of $k$-dependent field redefinitions is not restricted
to composite fields. We demonstrate this here by a transformation
of the gauge field, which becomes a $k$-dependent nonlinear combination
according to 
\begin{equation}\label{5.1}
\partial_t A_\mu(q)=-\partial_t\gamma_k(\bar\psi\gamma_\mu\psi)(q)
-\partial_t\delta_k(\partial_\nu F_{\mu\nu})(q)
-\partial_t\zeta_k(\partial_\mu\partial_\nu A_\nu)(q).
\end{equation}
This transformation
can absorb the vector channel in the four-fermion interaction.
Indeed, we may enlarge our truncation (2) by a term
\begin{equation}\label{5.2}
\Gamma_k^{(V)}=\int d^4x\bar\lambda_{v,k}(\bar\psi\gamma_\mu\psi)
(\bar\psi\gamma_\mu\psi)
\end{equation}
(or a corresponding generalization with momentum-dependent
coupling $\bar\lambda_{v,k}$). The flow equation
for $\bar\lambda_v$ reads
\begin{equation}\label{5.3}
\partial_t\bar\lambda_{v,k}=-6k^{-2}v_4l_{1,2}^{(FB)4}(0,0)e^4
+\frac{1}{2}k^{-2}v_4\frac{1}{Z^2_{\phi,k}}l^{(FB)4}_{1,2}
(0,\case{\bar m^2_k}{Z_{\phi,k}k^2})\bar h^4_k
+e\partial_t\gamma_k+{\cal O}(\bar\lambda_{\sigma,k},\bar\lambda_{v,k}).
\end{equation}
In the following, we again omit the term $\sim \bar h_k^4$, whose
contributions are subdominant once the scalars have decoupled from the
flow. Choosing $\gamma_k$ according to
\begin{equation}\label{5.4}
\partial_t\gamma_k=6k^{-2}v_4l^{(FB)4}_{1,2}(0,0)e^3,
\end{equation}
we can obtain a vanishing of $\bar\lambda_v$ for all $k$.
This procedure introduces additional terms
$\sim \bar\sigma_k(\partial_\nu F_{\mu\nu})\bar\psi\gamma_\mu\psi$
with $\bar\sigma_k$ obeying
\begin{equation}\label{35}
\partial_t\bar\sigma_k=-\partial_t\gamma_k+e\partial_t\delta_k+\dots,
\end{equation}
where the dots correspond to contributions from $\partial_t\Gamma$
at fixed fields. Adjusting $\delta_k$ permits us to enforce
$\bar\sigma_k=0$. As a result, only the gauge field propagator 
gets modified by higher derivative terms. We note that the modified
gauge field has the same gauge transformation properties as the original
field only for $\zeta_k=0$. In fact, the gauge fixing becomes 
dependent on the fermions by a term $\bar\sigma_k^{(gf)}\bar\psi
\gamma_\mu\psi\partial_\mu\partial_\nu A_\nu$ according to
\begin{equation}\label{36} 
\partial_t\bar\sigma_k^{(gf)}=\frac{1}{\alpha}\partial_t\gamma_k
+e\partial_t\zeta_k+\dots
\end{equation}
Again, we can enforce a vanishing $\bar\sigma_k^{(gf)}$ for all
$k$ by an appropriate choice of $\zeta_k$. The contribution to
the evolution of the gauge field propagator resulting from the
field redefinition (\ref{5.1}) is 
\begin{equation}\label{37}
\partial_t\Gamma^{(A2)}=-\partial_t\delta_k(\partial_\nu
F_{\mu\nu})(\partial_
\rho F_{\mu\rho})+\frac{1}{\alpha}\partial_t\zeta_k
(\partial_\mu\partial_\nu A_\nu)(\partial_\mu\partial_\rho A_\rho)
+\dots 
\end{equation}
With $\partial_t\gamma_k\sim e^3,\ \partial_t\delta_k\sim e^2,\
\partial_t\zeta_k\sim e^2/\alpha$ we see that the field  
redefinitions lead to a modification of the kinetic term (or a
momentum-dependent wave function renormalization 
of the gauge field) already in leading order $\sim e^2$. Depending
on the precise definition of the renormalized gauge coupling this can 
modify the $\beta$-function for the ``composite gauge field''
as compared to the original one. This modification is the counterpart
of the elimination of the effective vertices $\sim \bar\sigma_k,
\bar\sigma_k^{(gf)}$. (We note that no corrections arise if $e$ is
defined by the effective electromagnetic vertex at very small
momentum.) 

\section{Conclusions}

It is an inherent feature of quantum field theory that a system with
certain fundamental degrees of freedom at a ``microscopic'' scale can
exhibit completely different degrees of freedom at a ``macroscopic''
scale, which appear to be equivalently ``fundamental'' in an
operational sense. A prominent example are the pions in a low-momentum
effective theory for strong interactions. These different faces of one
and the same system are related by the action of the renormalization
group. In the present work, we realize this formal concept with the
aid of a renormalization group flow equation for the effective average
action whose field variables are allowed to change continuously under
the flow from one scale to another. In particular, this generally
nonlinear transformation of variables is suitable for studying the
renormalization flow of bound states.

We illustrate these ideas by way of example by considering the gauged
NJL model at weak gauge coupling. Our flow equations can clearly
identify the phase transition to spontaneous chiral symmetry breaking.
In our picture, the interaction between the fermions, representing the
fundamental degrees of freedom at high momentum scales, gives rise to
a pairing into scalar degrees of freedom. These so-formed bound states
may still appear effectively as composite objects at lower scales or
rather as fundamental degrees of freedom, depending on the strength of
the initial interaction. As the criterion that distinguishes between
these two cases, we classify the renormalization flow of the scalar
bound states: ``fundamental behavior'' is governed by a typical
infrared unstable fixed point with the relevant parameter
corresponding to the mass of the scalar. Contrary to this,
``bound-state behaviour'' is related to an infrared attractive
(partial) fixed point that is governed by the relevant and marginal
parameters of the ``fundamental'' fermion and photon -- massless QED
in our case. The flow may show a crossover from one to the other
characteristic behavior. This physical picture is obtained from the
continuous transformation of the field variables under the flow that
translates the fermion interactions into the parameters of the scalar
sector. In the case of spontaneous chiral symmetry breaking, the
scalars always appear as ``fundamental'' on scales characteristic for
the phase transition and the order parameter.

From a different perspective, we propose a technique for performing a
bosonization of self-interactions of fundamental fermion fields
permanently at all scales during the renormalization flow. Provided
that appropriate low-energy degrees of freedom of a quantum system are
known, our modified flow equation for the average effective action is
capable of describing the crossover from one set of variables to
another during the flow in a well-controlled manner. Thereby, the
notions of fundamental particle or bound state become scale dependent.

For the translation from fermion bilinears to scalars, the gauge field
acts rather as a spectator, permanently catalyzing the generation of
four-fermion interactions under the flow. In the vector channel,
however, the gauge field can also participate in the field
transformation. Hereby, a four-fermion interaction $\sim
(\yb\gamma^\mu\psi)^2$ is absorbed at the expense of a modified photon
kinetic term, which can lead to a change in the beta function
$\beta_e$ of an appropriately defined effective gauge coupling. We
expect this type of transformation to be particularly useful in the
strong-gauge-coupling sector of the gauged NJL model. Here it is known
that the four-fermion interaction can acquire an anomalous scaling
dimension of 4 (instead of 6) \cite{Leung:1986sn}, so that it mixes
with the gauge interaction (in a renormalization-group sense) anyway.
It should be worthwhile to employ this transformation for a search for
the existence of ultraviolet stable fixed points in the $\beta_e$
function, to be expected for a large number of fermion species
$N_{\text{F}}$ \cite{Reenders:2000bg}.

In view of the motivating cases of top quark condensation in the Higgs
sector and color octet condensation in low-energy QCD, we now have an
important tool at our disposal which allows for a nonperturbative
study of the transition from the underlying theory to the condensing
degrees of freedom. Particularly in the case of ``spontaneous breaking
of color'', a quantitatively reliable calculation of the potential for
the quark-antiquark degrees of freedom seems possible. Analogously to
the gauged NJL model, the effective quark self-interactions, being
induced by the exchange of gluons and instantons, have to be
translated into the scalar bound-state sector. The renormalization
flow of the latter and the symmetry properties of their corresponding
potential shall finally adjudicate on ``spontaneous breaking of
color''.

\section*{Appendices}

\renewcommand{\thesection}{\mbox{\Alph{section}}}
\renewcommand{\theequation}{\mbox{\Alph{section}.\arabic{equation}}}
\setcounter{section}{0}
\setcounter{equation}{0}

\section{Dirac algebra and Fierz transformations}

We work in a chiral basis, $\psi=\left( \begin{array}{c} \yl\\
    \yr\end{array} \right)$, $\yb=(\ybr,\ybl)$, where $\psi$ and $\yb$ 
 are anticommuting Grassmann variables and should be considered as
independent, $\psi_L=\frac{1}{2}(1+\gamma_5)\psi$.
The Dirac algebra for 4-dimensional Euclidean spacetime is given
by 
\begin{eqnarray}
\{ \gamma_\mu,\gamma_\nu \}&=&2 \delta_{\mu\nu}
,  \quad 
\gamma_\mu =(\gamma_\mu)^\dagger,\nonumber
 \\
\gamma_\mu\gamma_\nu&=&\delta_{\mu\nu} -\I \sigma_{\mu\nu}, \quad
\sigma_{\mu\nu}=\case{\I}{2} [\gamma_\mu,\gamma_\nu],  \nonumber\\
\gamma_5&=&\gamma_1\gamma_2\gamma_3\gamma_0.\label{1.2}
\end{eqnarray}
Defining $
O_{\text{S}}=\mathbbm{1},\, O_{\text{V}}=\gamma_\mu,\, O_{\text{T}}
=\frac{1}{\sqrt{2}} \sigma_{\mu\nu}, \,
O_{\text{A}}=\I\gamma_\mu\gamma_5 ,\, O_{\text{P}}=\gamma_5$,
we obtain the Fierz identities in the form
\begin{equation}
(\yb^a O_X\psi_b)(\yb_c O_X \psi_d) =\sum_Y C_{XY} (\yb_a O_Y
\psi_d)(\yb_c O_Y \psi_b), \label{F6}
\end{equation}
where $X,Y=$S,V,T,A,P and
\begin{equation}
C_{XY}=\left( \begin{array}{ccccc}
-\case{1}{4} & -\case{1}{4} & -\case{1}{4} & -\case{1}{4}& -\case{1}{4} \\
-1           & \case{1}{2}  &      0       & -\case{1}{2}& 1 \\
-\case{3}{2} &      0       & \case{1}{2}  &      0      & -\case{3}{2} \\
-1           & -\case{1}{2} &      0       & \case{1}{2} &  1 \\ 
-\case{1}{4} & \case{1}{4}  & -\case{1}{4} & \case{1}{4} & -\case{1}{4} 
\end{array} \right).\label{F7}
\end{equation}
The structure  $(\yb O_{\text{V}} \psi)^2-(\yb O_{\text{A}} \psi)^2$ is
invariant under Fierz transformations, and $(\yb O_{\text{V}}
\psi)^2+(\yb O_{\text{A}} \psi)^2$ can be completely transformed into
(pseudo-)scalar channels:
\begin{equation}
(\yb O_{\text{V}} \psi)^2+(\yb O_{\text{A}} \psi)^2= -2[ (\yb
O_{\text{S}} \psi)^2-(\yb O_{\text{P}} \psi)^2 ].\label{new2}
\end{equation}
Further useful identities are
\begin{eqnarray}
(\yb O_{\text{T}}\gamma_5 \psi)^2&=&(\yb O_{\text{T}} \psi)^2,
  \nonumber\\
(\yb \gamma_\alpha\gamma_\beta\gamma_\delta\psi) 
  (\yb \gamma_\alpha\gamma_\beta\gamma_\delta\psi) &=&10
  (\yb\gamma_\mu\psi)^2 +6(\yb\gamma_\mu \gamma_5\psi)^2, \nonumber\\
(\yb \gamma_\alpha\gamma_\beta\gamma_\delta\psi) 
  (\yb \gamma_\delta\gamma_\beta\gamma_\alpha\psi) &=&10
  (\yb\gamma_\mu\psi)^2 -6(\yb\gamma_\mu \gamma_5\psi)^2. \label{new3}
\end{eqnarray}

\section{Details of the fluctuation matrix}
\setcounter{equation}{0}

In Eq.\re{5}, we decompose the fluctuation matrix into ${\cal P}$ and
${\cal F}$, the latter containing the field dependence. The inverse
propagator is diagonal in momentum space,
\begin{equation}
{\cal P}\!=\!\! \left(\!\! \begin{array}{ccccc}
q^2(1+r_B)  & & & & \\
& 0& Z_{\phi,k}q^2(1+\!r_B)+\mkq & & \\
& Z_{\phi,k}q^2(1+\!r_B)+\mkq & 0 & & \\
& &  & 0 & -\fss{q}^T(1+r_{\text{F}}) \\
& &  & -\fss{q}(1+r_{\text{F}}) & 0 \end{array}\!\! \right)\!.\label{B1}
\end{equation}
It involves the dimensionless cutoff functions $r_B$ and
$r_{\text{F}}$, being related to the components of $R_k$ by
\begin{equation}
R_k^A=q^2 r_B,\quad R_k^\phi=Z_{\phi,k} q^2 r_B, \quad R_k^\psi=-\fss{q}\,
r_{\text{F}}. \label{B2}
\end{equation}
Of course, these cutoff functions are supposed to satisfy the usual
requirements of cutting of the infrared and suppressing the
ultraviolet sufficiently strongly.  The conventions for the Fourier
transformation employed here can be characterized by
\begin{equation}
\psi(x)=\int \frac{d^4q}{(2\pi)^4}\, \E^{\I qx}\, \psi(q), \quad
\yb(x)=\int \frac{d^4q}{(2\pi)^4}\, \E^{-\I qx}\, \yb(q) \label{B1a}
\end{equation}
for the fermions. As a consequence, the Fourier modes of the field
$\Phi$ and $\Phi^T$ in Eq.\re{new4} are then given by
$\Phi(q)=(A(q),\phi(q),\phid(-q),\psi(q),\yb^T(-q))$ (column vector)
and $\Phi^T(-q)=(A^T(-q),\phi(-q),\phid(q),\psi^T(-q),\yb(q))$ (row
vector). Owing to the sign difference in the arguments of $\psi$ and
$\yb$, the inverse propagator ${\cal P}$ in Eq.\re{B1} is symmetric
under transposition $T$. Concerning the field-dependent part, the
matrix $\cal F$ is also diagonal in momentum space for constant
``background'' fields and antisymmetric under transposition in all
fermion-related components:
\begin{eqnarray}
{\cal F}&=& \left( \begin{array}{ccccc}
0 & 0 & 0 &-e\yb\gamma_\mu & e\psi^T\gamma^T_\mu \\
0 & 0 & 0 & \hk\ybr & -\hk \yl^T \\
0 & 0 & 0 & -\hk\ybl&   \hk\yr^T \\
e\gamma^{T}_\mu\yb^T & -\hk\ybr^T & \hk\ybl^T  & \bar{H} & -F^T \\
-e\gamma_\mu\psi& \hk\yl & -\hk\yr  & F & H \end{array} \right),\label{B3}
\end{eqnarray}
where
\begin{eqnarray}
H&=&-\lk \bigl[ \psi\psi^T -\gamma_5\psi\psi^T\gamma_5\bigr],
  \quad H^T=-H, \nonumber\\
\bar{H}&=&-\lk \bigl[ \yb^T\yb -\gamma_5\yb^T\yb\gamma_5\bigr],
  \quad \bar{H}^T=-\bar{H}, \label{B4}\\
F&=&\hk(\pl\phi-\pr\phid) +\lk\bigl[
  (\yb\psi)-\gamma_5(\yb\gamma_5\psi)
  +\psi\yb-\gamma_5\psi\yb\gamma_5\bigr], \nonumber
\end{eqnarray}
and $\gamma_\mu^T$ is understood as transposition
in Lorentz and/or Dirac space. The projectors
$\pl$ and $\pr$ are defined as $P_{\text{L,R}}=(1/2)(1\pm\gamma_5)$.
In Eq.\re{B4}, we have dropped the $A_\mu$ dependence of the quantity
$F$ which is not needed for our computation.

\section{Exact flow equation for flowing field variables}
\label{flowvar}
\setcounter{equation}{0}

In the standard formulation of the flow equation
\cite{Wetterich:1993yh}, the field variables of the $k$-dependent
effective action $\Gamma_k[\phi]$ correspond to the so-called
classical field defined via
\begin{equation}
\phi=\frac{\delta W_k[j]}{\delta j} \equiv \phi_\Lambda, \label{ffv1}
\end{equation}
where all explicit $k$ dependence is contained in the cutoff
dependence of $W_k$, the generating functional for connected Green's
functions. The last identity in Eq.\re{ffv1} symbolizes that no
explicit $k$ dependence occurs for this classical field, and thereby
the field $\phi$ at any scale is identical to the one at the
ultraviolet cutoff $\Lambda$. (The functional dependence of $\phi$ on
$j$ contains, of course, an implicit $k$ dependence.)

In the present work, we would like to study the flow of the effective
action, now depending on a field variable that is allowed to vary
during the flow. For an infinitesimal change
of $k$, $\phi_k$ also varies infinitesimally:
\begin{equation}
\phi_{k-dk}(q)=\phi_k(q)+\delta\alpha_k(q)\,F[\phi_k,\dots](q),
\quad\partial_k\phi_k=-\partial_k\alpha_k\, F[\phi_k,\dots], \label{ffv4}
\end{equation}
where $\delta\alpha_k$ is infinitesimal and $F$ denotes some functional of
possibly all fields of the system. The desired effective action
$\Gamma_k[\phi_k]$ is derived from a modified 
functional $W_k$:
\begin{equation}
\E^{W_k[j,\dots]}=\int {\cal D}\chi\,{\cal D}(\dots)\, \E^{-S[\chi]
  -\Delta S_k[\chi_k] +\int j \chi_k +\dots}. \label{ffv5}
\end{equation}
The dots again indicate the contributions of further fields,
suppressed in the following, and we assume the quantum field $\chi$ to
be a real scalar for simplicity. In contrast to the common formulation
\cite{Wetterich:1993yh} the source $j$ multiplies a $k$-dependent
nonlinear field combination $\chi_k$ which obeys
\begin{equation}\label{C3A}
\partial_k\chi_k=-\partial_k\alpha_k\,G[\chi_k,...].
\end{equation}

We also modify the infrared cutoff
\begin{equation}
\Delta S_k[\chi_k]=\frac{1}{2} \int \chi_k R_k \chi_k, \label{ffv2}
\end{equation}
which ensures that the momentum modes $\sim k$ of the actual field
$\chi_k$ contribute to the flow at the scale $k$, regardless of its
different form at other scales.  Furthermore, the cutoff form of
Eq.\re{ffv2} shall lead us to a simple form of the flow equation. The
$k$-dependent classical field is given by
\begin{equation}
\phi_k:=\langle \chi_k\rangle= \frac{\delta W_k}{\delta j},
\label{ffv6}
\end{equation}
and, as a consequence, the higher derivatives of $W_k[j]$ are now
related to correlation functions of $\chi_k$ and no longer of
$\chi_\Lambda$.\footnote{ Eq.\re{ffv6} implies the
  relation $F[\phi_k,\dots]=\langle G[\chi_k,\dots]\rangle$. However,
  the definition of $\chi_k$ is often not needed explicitly. For our
purposes it suffices to define $F[\phi_k,\dots]$.} The desired
effective action is finally defined in the usual way via a Legendre
transformation including a subtraction of the cutoff:
\begin{equation}
\Gamma_k[\phi_k]=-W_k\bigl[j[\phi_k]\bigr] +\int j[\phi_k]\, \phi_k
-\Delta S_k[\phi_k]. \label{ffv7}
\end{equation}
Its flow equation is obtained by taking a derivative with respect to 
the RG scale $k$,
\begin{eqnarray}
\partial_k \Gamma_k[\phi_k] &=&\partial_k\Gamma_k[\phi_k]\bigl|_{\phi_k} +
  \int\frac{\delta \Gamma_k[\phi_k]}{\delta \phi_k} \, \partial_k \phi_k
  \nonumber\\
&=&\frac{1}{2} \Tr\,\frac{\partial_k R_k}{\Gamma_k^{(2)}[\phi_k] +R_k}
  -\int\frac{\delta\Gamma_k[\phi_k]}{\delta \phi_k} \, F[\phi_k,\dots]\,
  \partial_k\alpha_k. \label{ffv12}
\end{eqnarray}
The first term of this flow equation is evaluated for fixed $\phi_k$
and hence leads to the form of the standard flow equation with
$\phi_\Lambda$ replaced by $\phi_k$; the second term describes the
contribution arising from the variation of the field variable under
the flow. 

\noindent
Some comments should be made:

\noindent
1) The variation\re{ffv4} of the field during the flow is a priori
arbitrary; therefore, Eq.\re{ffv12} (together with some boundary
conditions) determines $\Gamma_k[\phi_k]$ completely only if
$\alpha_k$ is fixed.

\noindent
2) This redundancy can be used to arrive at a simple form for
$\Gamma_k[\phi_k]$ adapted to the problem under consideration.  For
example, one may determine $\alpha_k$ (and $F[\phi_k,\dots]$) in such
a way that some unwanted coupling vanishes.

\noindent
3) This program can be generalized straightforwardly to a whole set of
transformations $\alpha_k^i$ for different fields $i$.  Furthermore,
the whole functional dependence may be $k$ dependent by replacing
$\partial_k\phi^i_k=-\partial_k\alpha_k^iF^i\to -\hat{\cal F}_k^i$.

\noindent
4) The generating functional of $\phi_\Lambda$ 1PI Green's functions
$\Gamma_{k=0}[\phi_\Lambda]$ can be obtained from
$\Gamma_{k=0}[\phi_{k=0}]$ by choosing $\alpha_{k=0}=0$.  In practice,
however, it is often more convenient to use ``macroscopic degrees of
freedom'' $\phi_{k=0}$ different from the ``microscopic'' ones
$\phi_\Lambda$. Their respective relation then needs the computation
of the flow of $\alpha_k$.

\noindent
5) The present definition of the average action $\Gamma_k[\phi_k]$ is
different from the effective action $\Gamma_k[\hat{\phi}_k]$ that is
obtained by a field transformation of the flow equation with fixed
fields as described in appendix~\ref{fieldtrafo}.  More precisely,
consider the flow of the effective action $\Gamma_k[\phi_\Lambda]$ for
fixed $\phi_\Lambda$ and perform a finite $k$-dependent field
transformation $\hat{\phi}_k=\phih_k[\phi,\alpha_k]$; then, even if
the transformation was chosen in such a way that $\phih_k$ were
identical with $\phi_k$ of the present method, these effective actions
would not coincide. The cutoff term acts differently in the two cases.
In the case of a field transformation, the cutoff involves
$\phi_\Lambda$, which is subsequently expressed in terms of the new
variables, whereas, in the present case, the cutoff is readjusted at
each scale and involves $\chi_k$.  Although this does not affect
physical results for exact solutions of the flow, this might lead to
differences in approximate solutions of the flow, even if the
approximation is implemented in the same way in either case.

\section{Fermion-boson translation by field transformations
with fixed cutoff}
\label{fieldtrafo}
\setcounter{equation}{0}

Here, we shall present a third approach to fermion-boson translation
relying on the standard formulation of the flow equation in addition
to a finite field transformation. We intend to identify a field
transformation of the type
\begin{eqnarray}
\phih&=&\phi+\alb\ybl\yr \quad \Longleftrightarrow \quad
\phi=\phih-\alb\ybl\yr, \nonumber\\
\phihd&=&\phid-\alb\ybr\yl \quad \Longleftrightarrow\quad
\phid=\phihd+\alb\ybr\yl, \label{d1}
\end{eqnarray}
so that an appropriate choice of a finite $\alb$ can transform the
four-fermion coupling to zero. For simplicity, we work in the limit of
a point-like interaction and dispense with an additional
transformation of the type $\sim\beta_k\, \phi_k$. Within these
restrictions, we shall not find the physical infrared behavior
described in Sects.~\ref{conti} and \ref{rencoup}. The present study
is intended only for a quantitative comparison of the different
approaches, which can be done by restricting the field redefinitions
in Sect.~\ref{conti} to Eq.\re{ffv13} with $q$-independent $\alpha_k$.

In contrast to the modified flow equation of Sect.~\ref{conti} and
appendix \ref{flowvar}, the source term and the infrared cutoff
considered here involve the original fields. This approach therefore
corresponds simply to a variable transformation in a given
differential equation (exact flow equation). The transformed effective
action for the hatted fields is obtained by simple insertion,
$\Gamma_k[\phih,\psi,A]:=\Gamma_k[\phi[\phih],\psi,A]$.  Except for
additional derivative terms arising from the scalar kinetic term, the
two actions are formally equivalent, where the new ``hatted''
couplings read in terms of the original ones
\begin{eqnarray}
\hmkq&=&\mkq,\nonumber\\
\hhk&=&\hk+\mkq\alb,\label{d3}\\
\hlk&=&\lk-\hk\alb-\case{1}{2} \mkq\alb^2.\nonumber
\end{eqnarray}
Again, the transformation function $\alb$ is finally fixed by
demanding that the beta function $\hat{\beta}_{\lambda_\sigma}$ for
the hatted four-fermion coupling $\hlk$ vanishes,
\begin{equation} 
\hat{\beta}_{\lambda_\sigma}(\hmkq,\hhk,\hlk,\alb,\pat\alb)=0,
\label{d6}
\end{equation}
with the boundary conditions $\bar{\lambda}_{\sigma,k=\kB}=0$
and $\bar{\alpha}_{k=\kB}=0$, which express complete
bosonization at $\kB$ (this also implies
$\hat{\lambda}_{\sigma,k=\kB}=0$). The new beta functions can
now be determined from the standard flow equation, being subject to
the field transformation. Following appendix A of
\cite{Wetterich:1996kf}, the basic equation is
\begin{eqnarray}
\pat\Gamma_k[\Phih]&=&\pat\Gamma_k\bigr|_\Phi
-\pat\phihd\bigr|_\Phi\frac{\delta}{\delta\phihd}\Gamma_k[\Phih]
 -\pat\phih\bigr|_\Phi \frac{\delta}{\delta\phih}\Gamma_k[\Phih]
 \nonumber\\
&=&\frac{1}{2}\, \STr\, \patt\,\ln\bigl(\Gamma_k^{(2)}+R_k\bigr) -\alb
\left[\ybl\yr \frac{\delta\Gamma_k}{\delta\phih}-\ybr\yl
  \frac{\delta\Gamma_k}{\delta\phihd} \right]. \label{d9}
\end{eqnarray}
Although there seems to be a formal resemblance to Eq.\re{ffv16},
there is an important difference: Eq.\re{d9} is equivalent to the
standard flow equation, whereas Eq.\re{ffv16} is not; the latter is
derived with a different cutoff term! Without resorting to the
calculation of Sect.~\ref{flowequation}, we can evaluate this equation
completely from the transformed truncation $\Gamma_k[\phih,\psi,A]$
and the field transformations\re{d1} according to
\begin{eqnarray}
\bigl(\Gamma_k^{(2)}\bigr)^T_{ab}&\equiv&
\frac{\overrightarrow{\delta}}{\delta \Phi^T_a}
 \Gk \frac{\overleftarrow{\delta}}{\delta \Phi_b} 
 \label{d10}\\
&=& 
  \left( \frac{\overrightarrow{\delta}}{\delta\Phi^T_a}\Phih^T_i\right) 
   \frac{\overrightarrow{\delta}}{\delta \Phih^T_i}
  \, \Gamma_k\,  \frac{\overleftarrow{\delta}}{\delta\Phih_j}
 \left(\Phih_j \frac{\overleftarrow{\delta}}{\delta\Phi_b}\right) 
+ (-1)^{(\Phih,\Phi^T)}
  \left(\Gamma_k\,\frac{\overleftarrow{\delta}}{\delta\Phih_i}\right)
  \left( \frac{\overrightarrow{\delta}}{\delta\Phi^T_a}\, \Phih_i\,
  \frac{\overleftarrow{\delta}}{\delta\Phi_b}\right),
  \nonumber
\end{eqnarray}
where $(\Phi_l,\Phi_m)=1$ iff fermionic components in $\Phi_l$ as well
as $\Phi_m$ are considered, and $(\Phi_l,\Phi_m)=0$ otherwise; the
indices $a,b,i,j$ label the different field components of $\Phi,\Phih$. 

From Eq.\re{d9}, or equivalently Eq.\re{d3}, we deduce that the
desired hatted beta functions are related to the original ones by
\begin{eqnarray}
\pat\hmkq\equiv\,\hat{\beta}_m&=&\beta_m, \nonumber\\
\pat\hhk\equiv\,\,\hat{\beta}_h&=&\beta_h+\alb
  \beta_m+\hmkq\pat\alb,\label{d34}\\ 
\pat\hlk\equiv\hat{\beta}_{\lambda_\sigma}&=&\beta_{\lambda_\sigma}-
  \alb \beta_h-\case{1}{2} \alb^2 \beta_m- \hhk\pat\alb, \nonumber
\end{eqnarray}
where the right-hand sides of Eq.\re{d34} have to be expressed in
terms of the hatted couplings by means of the relations\re{d3}. Now we
determine $\alb$ by demanding that $\hat{\beta}_{\lambda_\sigma}$
vanishes for vanishing $\hlk$, so that no four-fermion coupling arises
during the flow. Introducing dimensionless quantities for the hatted
couplings, $\talpha_k=k^2Z_{\phi,k}^{1/2} \alb$,
$\epsilon_k=k^{-2}Z_{\phi,k}^{-1} \hmkq$, $h_k=Z_{\phi,k}^{-1/2}\hhk$,
we end up with the flow equations
\begin{eqnarray}
\pat \epsilon_k&=&-2\epsilon_k+\frac{1}{8\pi^2}
   (h_k-\epsilon_k\talpha_k)^2, \nonumber\\
\pat( h_k-\epsilon_k
   \talpha_k)&=&\left[-\frac{e^2}{2\pi^2}-\frac{1}{4\pi^2} \talpha_k
   (h_k -\case{1}{2} \epsilon_k \talpha_k)\right](h_k -\epsilon_k 
  \talpha_k), \nonumber\\
\pat \talpha_k&=&2\talpha_k - \frac{9}{8\pi^2} \frac{e^4}{h_k} 
  -\frac{1}{2\pi^2}\, e^2\, \talpha_k 
   +\frac{1}{16\pi^2} (h_k-2\epsilon_k\talpha_k+\case{\epsilon_k^2}{2}
   \case{\talpha_k^2}{h_k} )\talpha_k \label{d41a}\\
&&+\frac{1}{8\pi^2}\frac{2+\epsilon_k}{
     (1+\epsilon_k)^2}\frac{1}{h_k}   
  (h_k-\case{1}{2} \epsilon_k\talpha_k)(h_k -\epsilon_k\talpha_k)^2\talpha_k
  \nonumber\\
&&+\frac{1}{32\pi^2} \frac{3 +\epsilon_k}{(1 +\epsilon_k)^3}
\frac{1}{h_k}(h_k-   \epsilon_k\talpha_k)^4, \nonumber
\end{eqnarray}
where we have inserted the threshold-function values as given in
appendix \ref{optimized} for illustrative purposes. These equations
have to be read side by side with Eqs.\re{B} and\re{E}. Contrary to
the latter, the present flow equations are completely coupled; in
particular, the flow for $\tilde{\alpha}_k$ is not disentangled as it
is in the case of Eqs.\re{B} and\re{E}. In the flow equation for the
mass, we again observe a critical mass-to-Yukawa-coupling ratio at the
bosonization scale, corresponding to the infrared unstable fixed point
$\te_1^\ast$ mentioned in Eq.\re{O.11}: from a numerical solution, we
find that $\te_\Lambda|_{\text{crit}}
=\epsilon_\Lambda/h_{\kB}^2|_{\text{crit}} \simeq \te_1^\ast$ is
hardly influenced by the $\hk^4$ term. The actual initial value of
this ratio at the bosonization scale with respect to
$\te_\Lambda|_{\text{crit}}$ hence determines whether the system flows
towards the phase with dynamical symmetry breaking or not.

\begin{figure}
\begin{picture}(160,45)
\put(0,0){
\epsfig{figure=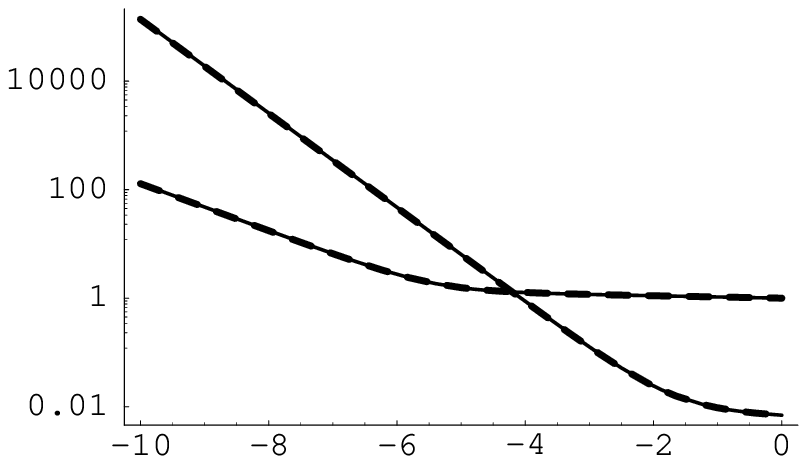,width=7cm}}
\put(22,38){$m_k^2$}
\put(19,17){$h_k$}
\put(66,0){$t$}
\put(81,0){
\epsfig{figure=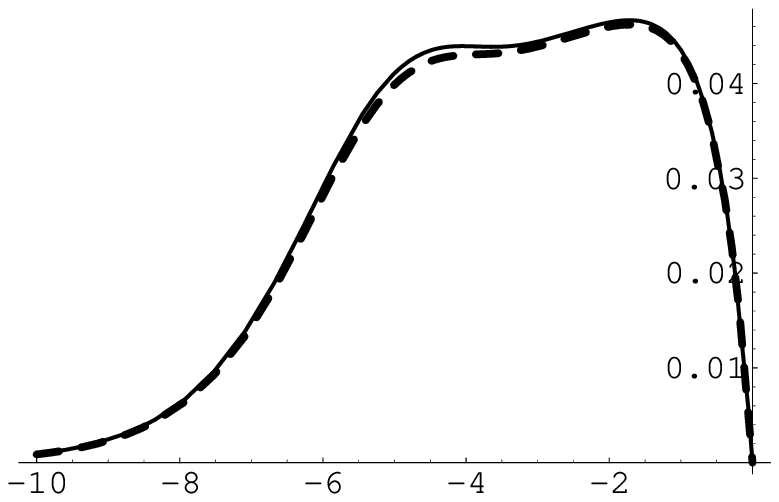,width=7cm}}
\put(143,45){$\alb$}
\put(145,0){$t$}
\end{picture} 
\caption{Flows of $\epsilon_k$, $h_k$ and $\tilde{\alpha}_k$ in the
  symmetric phase ($h_{\kB}=1$, $e=1$,
  $\epsilon_\Lambda=1.16\cdot[1/(16\pi^2)]$). The solid lines represent a
  solution to Eqs.\re{d41a}; the dashed lines correspond to the
  analogous flow employing the method of Sect.~\ref{conti} and
  appendix \ref{flowvar} (without the $\sim \pat\beta_k$
  transformation). The plots are representative for a wide range of
  initial conditions.}
\label{fig3} 
\end{figure}

In order to compare the present method with the one employed in
Sect.~\ref{conti}, we plot a numerical solution of Eqs.\re{d41a} in
Fig.~\ref{fig3} (solid lines) and compare it to a solution of the
corresponding equations\re{B} and\re{E} (dashed lines) without those
terms arising from the additional transformation $\sim\pat\beta_k$,
which is not considered in Eqs.\re{d41a}. In this figure, it becomes
apparent that both methods do not only agree qualitatively, but also
quantitatively to a high degree -- as they should. The minor differences
in these approaches can be attributed to the different formulation of
the cutoff, and thereby reflect the inherent cutoff dependence for
approximative solutions to the otherwise exact flow equation.

The same conclusion can be drawn from the flow equation for the
dimensionless combination $\te_k$ as defined in Eq.\re{CW8}. Although
the $\te_k$ flow equation derived from Eqs.\re{d41a} is comparably
extensive (we shall not write it down here) and not identical to
Eq.\re{CW9}, the fixed-point structure remains nevertheless the same,
and the $\te_k$ flow reduces exactly to Eq.\re{CW9} for $k\to\kB$,
where all our approaches agree. Moreover, the position of the infrared
stable fixed point $\te_2^\ast$ also remains the same in the infrared
to leading order in $e$, so that the different approaches describe the
same physics.

To summarize, employing the method of field transformation in the flow
equation for fixed cutoff, the same properties of the system can be
derived with a similar numerical accuracy in comparison to the flow
equation proposed in Sect.~\ref{conti} and appendix \ref{flowvar}.
However, the structure of the resulting flow equations derived in this
appendix appears to be more involved, and we expect this to be a
generic feature of field transformation in the flow equation for fixed
cutoff -- at least within the usual approximation schemes.

\section{Cutoff Functions}
\label{optimized}
\setcounter{equation}{0}

For concrete computations, we have to specify the cutoff functions.
Here we shall use optimized cutoff functions as proposed in
\cite{Litim:2001up}, which furnish a fast convergence behavior and
provide for simple analytical expressions. Employing the nomenclature
of \cite{Jungnickel:1996fp}, we use the dimensionless cutoff functions
($y=q^2/k^2$) 
\begin{eqnarray}
r_B(y)&=&\left(\frac{1}{y}-1 \right)\theta(1-y), \quad
  p(y)=y(1+r_B(y))=y+(1-y)\, \theta(1-y), \nonumber\\
r_{\text{F}}(y)&=&\left(\frac{1}{\sqrt{y}} -1\right) \theta(1-y),
  \quad p_{\text{F}}(y)=y(1+r_{\text{F}}(y))^2 \to p(y). \label{o1}
\end{eqnarray}
Here we have set the normalization constants $c_{\text{B}}$ and
$c_{\text{F}}$ mentioned in \cite{Litim:2001up} to the values
$c_{\text{B}}=1/2$ and $c_{\text{F}}=1/4$, so that fermionic and
bosonic fluctuations are cut off at the same momentum scale
$q^2=k^2$. This is natural in our case in order to avoid the
situation in which fermionic modes which are already integrated out are
transformed into bosonic modes which still have to be integrated out
or vice versa. 

For these cutoff functions, the required threshold functions evaluate
to 
\begin{eqnarray}
l^{(F)\,d}_n(\omega)&=&(\delta_{n,0}+n)\frac{2}{d}
\frac{1}{(1+\omega)^{n+1}}, \label{o6}\\
l^{(FB)\,d}_{n_1,n_2}(\omega_1,\omega_2)&=&\frac{2}{d}
\frac{1}{(1+\omega_1)^{n_1}(1+\omega_2)^{n_2}}\left[
  \frac{n_1}{1+\omega_1} +\frac{n_2}{1+\omega_2} \right]. \label{o9}
\end{eqnarray}

\section*{Acknowledgment}
H.G. would like to thank D.F.~Litim for discussions on optimized
cutoff functions and acknowledges financial support by the Deutsche
Forschungsgemeinschaft under contract Gi 328/1-1.

\end{document}